\documentclass[aps,twocolumn,prl,showpacs,showkeys,preprintnumbers,superscriptaddress,nobibnotes,floatfix,longbibliography,notitlepage,nofootinbib]{revtex4-2}
\usepackage{graphicx}
\usepackage{epstopdf}
\usepackage{amsmath}
\usepackage{amsfonts}
\usepackage{amssymb}
\usepackage{appendix}
\usepackage{enumerate}
\usepackage{natbib}
\usepackage{comment}
\usepackage{bbold}
\usepackage[shortlabels]{enumitem}
\usepackage{color}
\usepackage{slashed}
\usepackage{subfigure}
\usepackage{setspace}
\usepackage{footnote}
\usepackage{lipsum}
\usepackage{multirow}
\usepackage[colorlinks = true,
            linkcolor = blue,
            urlcolor  = blue,
            citecolor = blue,
            anchorcolor = blue]{hyperref}
\usepackage[capitalize]{cleveref}
\usepackage{braket}
\usepackage{multirow}
\usepackage{physics}
\usepackage[normalem]{ulem}
\usepackage{url}
\usepackage{units}
\usepackage[normalem]{ulem}
\usepackage{graphicx}
\usepackage{dcolumn}
\usepackage{bm}
\usepackage{lipsum}
\usepackage{lettrine}

\definecolor{pink}{rgb}{0.858, 0.188, 0.478}

\begin{document}

\preprint{USTC-ICTS/PCFT-24-07}
\bibliographystyle{apsrev4-1}
\title{
Supernovae Time Profiles as a Probe of New Physics at Neutrino Telescopes
}

\author{Jeffrey Lazar}
\email{jlazar@icecube.wisc.edu}
\affiliation{Centre for Cosmology, Particle Physics and Phenomenology -- CP3, Universit\'{e} catholique de Louvain, Louvain-la-Neuve, Belgium}
\author{Ying-Ying Li}
\email{liyingying@ihep.ac.cn}
\affiliation{Institute of High Energy Physics, Chinese Academy of Sciences, Beijing 100049, China}
\affiliation{Peng Huanwu Center for Fundamental Theory, Hefei, Anhui 230026, China}
\affiliation{Interdisciplinary Center for Theoretical Study, University of Science and Technology of China, Hefei, Anhui 230026, China}
\author{Carlos A. Arg\"{u}elles}
\email{carguelles@g.harvard.edu}
\affiliation{Department of Physics \& Laboratory for Particle Physics and Cosmology, Harvard University, Cambridge, MA 02138, USA}
\author{Vedran Brdar}
\email{vedran.brdar@okstate.edu}
\affiliation{Department of Physics, Oklahoma State University, Stillwater, OK, 74078, USA}


\begin{abstract}
Neutrino telescopes, including IceCube, can detect galactic supernova events by observing the collective rise in photomultiplier count rates with a sub-second time resolution. 
Leveraging precise timing, we demonstrate for the first time the ability of neutrino telescopes to explore new weakly coupled states emitted from supernovae and subsequently decaying to neutrinos.
Our approach utilizes publicly available packages, \texttt{ASTERIA} and \texttt{SNEWPY}, for simulating detector responses and parametrizing neutrino fluxes originating from Standard Model and new physics. 
We present results for two Beyond-the-Standard Model scenarios and introduce the tool developed for testing a diverse range of new physics models.
\end{abstract}

\maketitle

\textbf{\textit{Introduction}}---The Standard Model (SM) is a remarkable but incomplete theory. 
Puzzles such as the non-vanishing neutrino mass, the origin of observed matter-antimatter asymmetry, and the nature of dark matter, among others, seek explanations beyond the SM (BSM).
Since there are also no firmly established clues on the energy scale at which the new particles should appear, numerous BSM searches across a wide range of scales have been performed: from collider searches at high energies~\cite{Nath:2010zj} to studies of cosmic microwave background at temperatures near absolute zero~\cite{Baumann:2015rya}.

Supernovae (SNe) release most of their energy in neutrinos, offering unique opportunities to test BSM physics via neutrino interactions.
The duration of the burst~\cite{Kamiokande-II:1987idp,Bionta:1987qt,Baksan} and inferred total energy of neutrinos~\cite{Loredo:2001rx,Pagliaroli:2008ur,Huedepohl2010} from SN 1987A---the only SN observed in neutrinos---have already given tentative constraints on the energies that could be carried by light BSM particles produced in the interior of a star.
Indeed, many different BSM scenarios have been constrained in this way, see \textit{e.g.} Refs.~\cite{Raffelt:2011nc,Arguelles:2016uwb,Suliga:2020vpz,Lucente:2021hbp,Caputo:2022rca,Caputo:2021rux,PhysRevD.100.083002,DeRocco:2019njg,Kazanas:2014mca,Magill:2018jla}.
If, additionally, BSM particles emitted from SNe can decay to neutrinos en route to Earth, even stronger limits can be set, as shown, for instance, in Ref.~\cite{Fiorillo:2022cdq, Akita:2022etk} for Majoron-like bosons and in Ref.~\cite{Brdar:2023tmi} for a realization featuring a neutrino magnetic moment portal.
In contrast to SM neutrinos, BSM states produced in the SN core can freely stream out without further interactions.
Consequently, they exit with higher energies than neutrinos, typically around $\mathcal{O}(100)$~MeV.
Provided such particles decay to neutrinos, strong constraints can be set from the fact that neutrinos of $\mathcal{O}(100)$~MeV were not recorded from SN 1987A~\cite{Fiorillo:2022cdq, Brdar:2023tmi}.
Such strategies were utilized for neutrino experiments operating during the SN 1987A event.
Sensitivity projections for DUNE~\cite{DUNE:2015lol} and Hyper-Kamiokande~\cite{Hyper-Kamiokande:2018ofw} in light of anticipated galactic SN event were also performed. 

\begin{figure}[t!]
\vspace{0.27cm}
    \centering
    \includegraphics[width=0.48\textwidth]{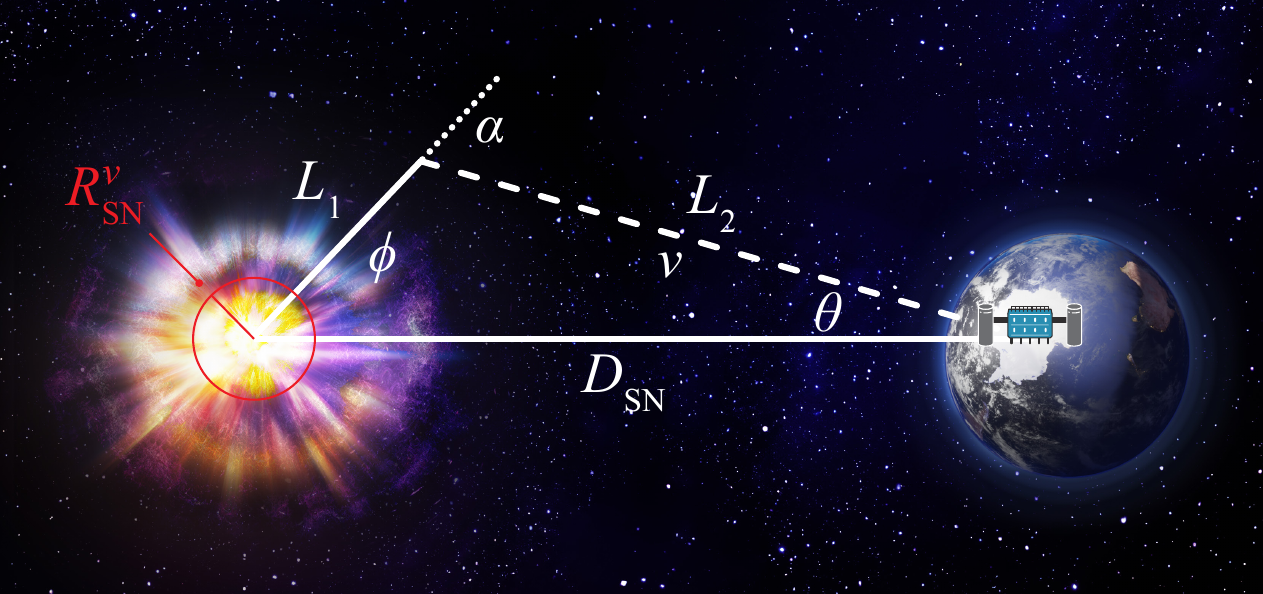}
    \caption{\textit{\textbf{The decay geometry for Majorons produced in a SN.}}
    Both the additional distance travelled relative to the direct line of sight and the potential non-relativistic speed of the Majoron can induce a delayed signal relative to the SM signal.
    }
    \label{fig:geometry}
\end{figure}

The IceCube Neutrino Observatory has pioneered the detection of neutrinos at $\mathcal{O}$(PeV) energies~\cite{IceCube:2014stg,IceCube:2018cha} and has identified several specific astrophysical neutrino sources~\cite{IceCube:2018cha,IceCube:2023ame,IceCube:2022der}.
IceCube is also expected to be able to detect the next galactic SN~\cite{Kopke_2011}.
In fact, a search for MeV neutrinos from optically obscured galactic SNe was recently performed in~\cite{IceCube:2023ogt} as well as the search for temporal correlation of MeV neutrino events with fast radio bursts~\cite{IceCube:2019acm}.
In both cases, the observable is a collective rise in all photomultiplier rates on top of the background noise in a certain time window, and it turns out that IceCube is sensitive to structures within an SN light curve as small as $\mathcal{O}(10^{-3})$ seconds. 

In fact, BSM states can also give rise to off-time signals in two ways.
First, if the BSM particles are non-relativistic, they will produce a delayed signal. 
However, it is also possible that new states are light, implying that they may even exit the SN at earlier times than certain flavors of SM neutrinos.
Such an early BSM signal, which to the best of our knowledge has not been previously considered, can, for instance, be induced by light BSM states produced from $\nu_e$ scattering inside the SN; they would then subsequently decay to $\bar{\nu}_e$, which chiefly induces the signal at IceCube. 
This would happen at early stages of an SN, around the neutralization burst, when $\nu_e$ are present, and $\bar{\nu}_e$ have not been produced yet.
Such novel timing patterns generically exist in models with new weakly interacting particles, such as the aforementioned Majoron model, the neutrino magnetic moment portal, and other BSM realizations \cite{Akita:2023iwq, Davoudiasl:2005fd}.

In this work, we will show that with the precise timing resolution, IceCube and other forthcoming neutrino telescopes will be able to provide powerful constraints on the BSM parameter space in association with forthcoming galactic SN events.\\


\textbf{\textit{Representative Model}}---As one representative case, we consider the Majoron model, studied in Ref.~\cite{Fiorillo:2022cdq}, where SN 1987A constraints were derived.
The relevant part of the Lagrangian reads
\begin{align}
\mathcal{L} \supset -g_{\alpha\beta} \nu_\alpha \nu_\beta \phi + \text{h.c.} - m_\phi \phi\phi^*\,,
\end{align}
where $\phi$ is the (pseudo)scalar, $m_\phi$ is its mass, and $g$ parametrizes interaction strength between $\phi$ and neutrinos.
We assume flavor universal interaction and hence $g_{\alpha\beta}\equiv g_\phi \, \delta_{\alpha\beta}$. 
Majorons are produced from (anti)neutrino coalescence in the star and subsequently decay to a pair of (anti)neutrinos, resulting in BSM (anti)neutrino flux emitted from the star.
The total decay width is given by $\Gamma_\phi = 3g^2 m_\phi/16 \pi$.

Using data from the simulation of $8.8 M_\odot$ progenitor star~\cite{Huedepohl2010} and including effects of neutrino oscillations assuming normal mass ordering~\cite{Dighe:1999bi}, we calculated standard neutrino fluxes at Earth.
Following Ref.~\cite{Fiorillo:2022cdq}, we also calculated the flux of emitted Majorons.
At $t \leq \unit[0.05]{s}$, Majorons are mainly produced through $\nu_e$ and $\nu_x$ coalescence where $\nu_x$ includes (anti)neutrinos of $\tau$ and $\mu$ flavor.
After $\unit[0.05]{s}$, the flux of Majorons decreases with the total flux of neutrinos of all flavors.
Thus, we found that the Majoron flux peaks around $t \sim \unit[0.05]{s}$.
We further cross-checked the agreement with Ref.~\cite{Fiorillo:2022cdq} by reproducing their bound from SN energy loss. 

Very weakly coupled Majorons immediately stream out after being produced inside the star.
On their way to Earth, Majorons of energy $E_\phi$ will travel at the speed of $\beta$ for a distance of $L_1$, then decay to (anti)neutrinos of energy $E_\nu$ at an emission angle $\cos\alpha = (2 E_\phi E_{\nu} - m^2_\phi)/(2E_{\nu}E_\phi\beta)$.
Daughter (anti)neutrinos will travel for a distance $L_2$ and reach Earth at the angle $\theta$ satisfying $D_{\rm SN} \sin\theta = L_1 \sin\alpha$ for a SN that is $D_{\rm SN}$ away; see the decay geometry in \cref{fig:geometry}.
The time delay, $\delta t$, relative to a massless particle moving along the line-of-sight path from a SN to reach the Earth is given by~\cite{Jaeckel:2017tud}
\begin{align}
    \delta t = \left(\frac{L_1}{\beta}-L_1\right) + (L_1 + L_2 -D_{\rm SN})\,,
    \label{eq:deltatexa}
\end{align}
where we have split the contribution to $\delta t$ into two parts: the first part comes from the slow-moving Majorons before their decay, and the second part is due to the detour from the line-of-sight path at the given emission angle $\alpha$. 

\begin{figure}[t!]
    \centering
    \includegraphics[width=0.47\textwidth]{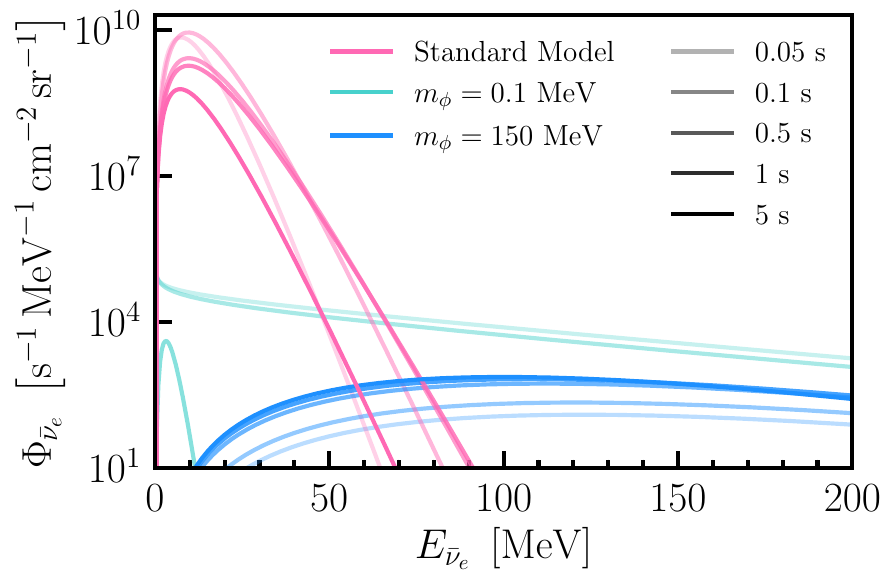}
    \caption{\textit{\textbf{Flux of $\bar{\nu}_{e}$ from SM production and two Majoron hypotheses at Earth}}.
    We choose the parameters $g_\phi = 10^{-9.25}$ and $g_\phi = 10^{-11.8}$ for $m_\phi = \unit[0.1]{MeV}$ and $m_\phi = \unit[150]{MeV}$, respectively.
    Only three lines appear for the light benchmark (teal lines), since after 0.5~s, the flux is negligible.
    }
    \label{fig:fluxes}
\end{figure}

We will consider a SN event that happens in the galaxy at a distance $D_{\rm SN}=\unit[10]{kpc}$, which is not unlikely~\cite{Reed:2005en,Rozwadowska:2020nab}. To get the characteristic value of $\delta t$, we can take $L_1 = L_\phi$ with $L_\phi = (E_\phi/m_\phi) \Gamma^{-1}_\phi \beta$ being the decay length of $\phi$.
As for the parameter regions we considered, $(D_{\rm SN} -D_{\rm SN} \cos\theta) \ll \unit[10^{-3}]{s}$, $\delta t$ can be approximated as
\begin{align}
    \delta t &\simeq L_\phi\left(\frac{1}{\beta} -1\right) + L_\phi(1-\cos\alpha) = \frac{8\pi}{3 E_{\nu}g_\phi^2}\,,
    \label{eq:deltat}
\end{align}
which is roughly independent of Majoron energy $E_\phi$.

\begin{figure}[t]
    \centering
    \includegraphics[width=0.475\textwidth]{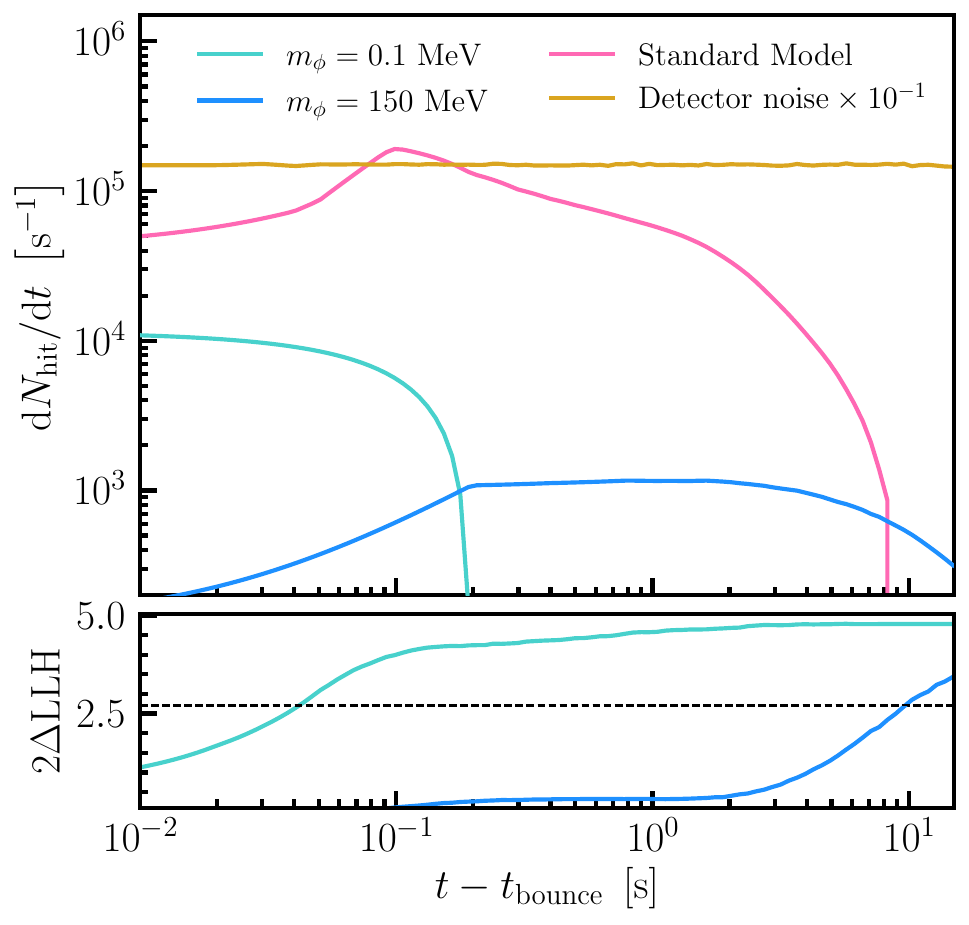}
    \caption{\textbf{\textit{Time profile of hits and test statistic.}}
    The top panel shows the number of hits as a function of the time after the SN bounce from the SM flux, two BSM scenarios, and detector backgrounds.
    In the bottom panel, we show the median of the SM-plus-BSM test statistic taken over 50,000 pseudodata realizations as a function of the same variable.
    The teal line corresponds to a BSM scenario in which the new state travels close to the speed of light, and thus, most of the test-statistic contribution comes before 0.1~s when the SM flux peaks.
    This should be contrasted with the sky blue line, which corresponds to a BSM scenario in which the new state travels subluminally.
    The dashed line shows the 95\textsuperscript{th} quantile of the SM-only test-statistic distributions.
    }
    \label{fig:hits_and_likelihood}
\end{figure}

We can obtain the flux of daughter $\nu$ by considering decays that occur only at $L_1\geq R^{\nu}_{\rm SN}$. As the smallest decay distance beyond which the daughter neutrino can escape the explosion unperturbed is roughly the radius of the core, we take $R^{\nu}_{\rm SN} = \unit[30]{km}$.
We will also limit our analysis to a maximal time window of $\unit[100]{s}$ after the SN bounce.
This time window is larger than typical time delay values arising in the model.

As $\bar{\nu}_e$ contribute chiefly to the signals at IceCube via inverse beta decay (IBD), we show in~\cref{fig:fluxes} the differential fluxes of daughter $\bar{\nu}_e$ at different times for two benchmark points in the Majoron model, in comparison to the SM $\bar{\nu}_e$ flux.
We expect the observed signal timing distribution at IceCube to follow these patterns of $\bar{\nu}_{e}$ flux. 

From \cref{fig:fluxes}, we can observe that the energy of $\bar{\nu}_e$ from Majoron decay extends to $\unit[100]{MeV}$ and above for both benchmarks, as expected. 
Consider the light Majoron case in \cref{fig:fluxes}.
Majorons, generating neutrinos with $E_{\bar{\nu}_e} \geq \unit[150]{MeV}$, are nearly relativistic with the emission angle $\alpha$ close to zero.
The time delay can be estimated from \cref{eq:deltat} as $\delta t \lesssim \unit[0.01]{s}$.
With such a negligible time delay, the time dependence of the $\bar{\nu}_e$ flux is mainly inherited from that of the Majorons upon production, with larger fluxes at $t\leq \unit[0.05]{s}$.
Notice that such a $\bar{\nu}_e$ flux would arrive at the detector earlier than the peak of the SM $\bar{\nu}_e$ flux around $t\simeq \unit[0.1]{s}$ (see \cref{fig:fluxes}), and it potentially leads to early signals at IceCube.
Following \cref{eq:deltat}, the time delay for $\bar{\nu}_e$ with energy $E_{\bar{\nu}_{e}} \sim \unit[10]{MeV}$ is typically $\gtrsim \unit[0.1]{s}$.
These $\bar{\nu}_e$ could be produced from Majorons with low energy of $\mathcal{O}(\unit[10]{MeV})$ or with high energies.
For the low-energy Majoron case, the time delay is mainly due to slow-moving Majorons, while for the high-energy case, due to the detour.
Such sizable time delays will shift these low energy $\bar{\nu}_e$ flux to later times, comparing to that of their mother Majorons, as exhibited in \cref{fig:fluxes}.

For the heavy Majoron case, the resulting $\bar{\nu}_e$ flux is larger at later times, manifesting the time delay from slowly moving Majorons.
For the heavy Majoron benchmark in \cref{fig:fluxes}, we can estimate that $\delta t \gtrsim \unit[10]{s}$ for $E_{\bar{\nu}_{e}}\sim \unit[200]{MeV}$ which is delayed compared to the peak of the SM $\bar{\nu}_e$ flux.
Compared to the light Majoron case, such a large time delay is a combination of a more slowly moving Majoron and a larger detour due to its large emission angle. As $L_1\lesssim L_\phi$, the time spread of the flux is roughly given by the aforementioned characteristic value of time delay, which is $\unit[0.1]{s}$ ($\gtrsim\unit[10]{s}$) for the light (heavy) benchmark case.

\textbf{\textit{Detector Response and Statistical Treatment}}---The IceCube Neutrino Observatory comprises 5,160 light-detecting digital optical modules (DOMs) buried in a cubic kilometer of the deep, transparent Antarctic ice sheet~\cite{IceCube:2016zyt}.
The DOMs are arranged on 86 strings of 60 DOMs each, with an interstring distance of 125~m (70~m for the eight centermost strings).
These DOMs detect the photons emitted by the charged byproducts of a neutrino interaction in or near the instrumented volume.
This enables IceCube to resolve individual neutrinos with energies ranging between a few GeV and a few PeV.
The neutrinos produced by SNe are far below this threshold and cannot be individually resolved; however, the immense number of neutrinos produced in an SN increases the detector's single-photon rate.
This dramatic increase in the rate of photons can be distinguished from the background caused by dark noise and radioactive activity in the DOM glass to enable the detection of galactic SN events~\cite{Griswold:2023iwz}.

Since more than 93\% of the photons detected in IceCube originate in IBD interactions~\cite{IceCube:2011cwc}, the SM-only event rate peaks at around 0.1~s.
BSM scenarios, on the other hand, can give rise to an early or delayed signal, depending on the model's particulars.
If light Majorons are created before the peak of the SM $\bar{\nu}_{e}$ flux, they would travel close to the speed of light and escape the SNe with negligible time delay.
They would subsequently decay to active neutrinos outside the SN, producing a flux of $\bar{\nu}_{e}$ that arrives before the SM flux.
On the other hand, higher-mass Majorons would travel sub-luminally and via larger deflections, producing a sizably delayed signal.

We use the \texttt{ASTERIA}~\cite{spencer_griswold_2020_3926835} package to simulate the detector response to the different SN scenarios and the thermal and radioactive noise in the DOMs.
This package simulates light yields from coherent $\nu_{e}$ scattering off electrons in the ice, IBD of $\bar{\nu}_{e}$ on nuclei in the ice, and charged- and neutral-current interactions for $\nu_{\alpha}$.
In addition to photons produced in neutrino interactions, it also simulates photons from thermal noise in the IceCube photomultipliers (PMTs) and from radioactive decay in the pressure glass housing.

We assess our sensitivity based on the test statistics $2\Delta $LLH.
For the SM-plus-BSM test statistics, we take the expected number of hits in each time bin $i$, given by:
\begin{equation}
\label{eq:mean}
\mu_{i} = \mu_{i}^{\mathrm{SM}} + \mu_{i}^{\mathrm{BSM}} + \mu_{i}^{\mathrm{BG}},
\end{equation}
where $\mu_{i}^{\mathrm{X}}$ is the expected number of hits in a particular bin from the SM, BSM, or detector backgrounds.
We generate pseudodata $d_i$ for each bin by sampling a number from a Poisson distribution with shape parameter $\mu_{i}$ from \cref{eq:mean}.
After defining the hypothetical model:
$$
m_{i}\left(\vec{n}\right) = n^{\mathrm{SM}}\mu_{i}^{\mathrm{SM}} + n^{\mathrm{BSM}}\mu_{i}^{\mathrm{BSM}} + n^{\mathrm{BG}}\mu_{i}^{\mathrm{BG}},
$$
where $\vec{n}=(n^{\mathrm{SM}},\, n^{\mathrm{BSM}},\, n^{\mathrm{BG}})$ are normalizations that globally scale all bins, we define the summed Poisson log-likelihood over time $t$ as:
\begin{equation}
    \mathcal{L}(\vec{n}, t ) = \sum_{i}^{\mathrm{bins}\leq t} d_{i}\log(m_{i}\left(\vec{n}\right)) - m_{i}\left(\vec{n}\right) - \log\left(d_{i}!\right).
\end{equation}
Then we fit the normalizations so as to maximize $\mathcal{L}(\vec{n}, \unit[100]{s})$ which is over the maximal time window $t = \unit[100]{s}$. For the SM-only hypothesis, we fit to $\vec{n}=(n^{\mathrm{SM}},\, 0,\, n^{\mathrm{BG}})$ which is maximized at $\vec{n}_{\rm SM}$, while for SM-plus-BSM hypothesis, we fit to $\vec{n}=(n^{\mathrm{SM}},\, n^{\mathrm{BSM}},\, n^{\mathrm{BG}})$ and it is maximized at $\vec{n}_{\rm BSM}$. 
With this, we can obtain the SM-plus-BSM test statistic at time $t$ as $2\Delta{\rm LLH} = 2\log\big(\mathcal{L}(\vec{n}_{\rm BSM}, t)/\mathcal{L}(\vec{n}_{\rm SM}, t)\big)$.
We then repeat this procedure many times---in practice on the order of 50,000---to create a distribution of test statistics for each model. The median of the test statistic values are shown in the bottom panel of \cref{fig:hits_and_likelihood} for two example BSM models (sky blue and teal lines).

Generating the pseudodata $d_i$ for each bin by sampling from a Poisson distribution with only the SM and BG shape parameter $\mu^{\rm SM}_{i}$ and $\mu^{\rm BG}_{i}$, we can similarly obtain the SM-only test statistic.
In \cref{fig:hits_and_likelihood}, we compare the median values of the SM-plus-BSM test statistics for two benchmark models---shown as the teal and sky blue lines---to the 95\textsuperscript{th} percentile of the SM-only test statistics, the dashed horizontal line\footnote{The SM-only distributions are nearly identical for both benchmark cases, and thus only one line is plotted.}.
The fact that the test statistics overtake the dashed line indicates that these BSM models should be detectable with 95\% confidence.

We observe that the main contribution to the SM-plus-BSM test statistic for the light benchmark comes from $t\lesssim \unit[0.1]{s}$ which is before the SM $\bar{\nu}_e$ peak, while for the heavy benchmark from $t \gtrsim \unit[3]{s}$ which is after the SM $\bar{\nu}_e$ peak.
As the distribution of test-statistic values is accumulated within a certain time interval for the 0.1 MeV benchmark, this analysis retains much of its sensitivity even with errors on the reconstruction of $t_{\mathrm{bounce}}\sim\mathcal{O}\left(10^{-3}~\mathrm{s}\right)$~\cite{Halzen_2009}.
Also note that as the SM-plus-BSM test statistic supersedes the SM-only test statistic at $t \lesssim \unit[10]{s}$, we don't expect uncertainties of SN modeling at a later time $t$ will significantly affect our results.

Furthermore, we define our exclusion sensitivity as the model at which the 5\textsuperscript{th} quantile of the SM-plus-BSM test-statistic distribution is greater than 0.
This corresponds to the model that, in the event of a null observation, we would exclude BSM with 95\% confidence.
In practice, we only simulate a finite number of points in the parameter space.
Thus, our sensitivity is the smallest coupling at which at least 95\% of the SM-plus-BSM test statistic is greater than 0.
This is a conservative choice, and given the granularity of our simulation, it results in an overestimation of the coupling of at most 25\%.

To estimate the sensitivity of IceCube Gen-2 to this model, we scale the expected number of hits by the effective photo-cathode area:
$$
A_{\mathrm{eff}}^{\mathrm{PC}} = \sum_{i} \varepsilon_{i} A_{i}^{\mathrm{PC}},
$$
where $\varepsilon_{i}$ and $A_{i}^{\mathrm{PC}}$ are the quantum efficiency and photocathode area of a PMT, and the sum runs over all PMTs in the detector.
We then carry out the same statistical procedure with these new distributions.


\textbf{\textit{Results and Discussions}}---
\begin{figure}[t!]
    \centering
    \includegraphics[width=0.47\textwidth]{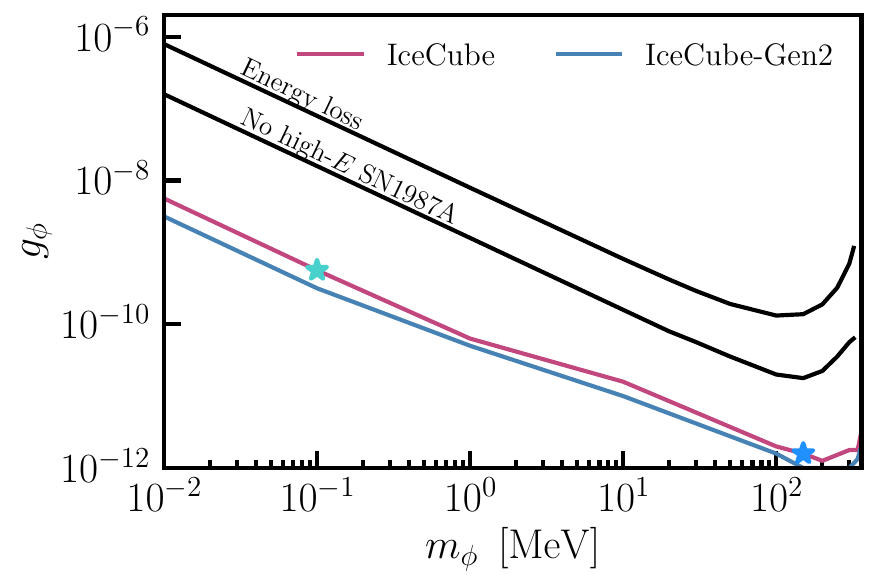}
    \caption{\textbf{\textit{Exclusion sensitivities for the Majoron case.}}
    The lines on this plot show the exclusion sensitivity for IceCube and IceCube Gen-2 at $2\sigma$ CL.
    Additionally, we also show the reaches by SN cooling and by the lack of observation of high-energy neutrinos from SN1987A as black lines. 
    The sensitivities shown by the dashed line are obtained following~\cite{Brdar:2023tmi} with time-integrated data.
    The “ceiling” of the SN cooling bound where $\phi$ could be trapped is unclear yet \cite{Fiorillo:2022cdq}.  
    }
    \label{fig:sensitivity}
\end{figure} 
In \cref{fig:sensitivity}, we show the expected exclusions that IceCube will set in the event that it does not see an off-time signal consistent with BSM physics.
To illustrate the broad physics application of the analysis, we also show results for the sterile neutrino case with magnetic portal in the Supplemental Materials.
Current limits~\cite{Fiorillo:2022cdq} from the energy loss requirement, and non-observation of high-energy SN neutrinos from SN 1987A are also shown in \cref{fig:sensitivity} for the Majoron case.
A future SN explosion would enable IceCube to improve current limits by one order of magnitude. 
Moreover, while current limits from time-integrated observables follow the behavior of $g_\phi\propto m^{-1}_\phi$ as Majoron production rate is proportional to $(g_\phi m_\phi)^2$, IceCube can provide a stronger limit than this by focusing on the time window, especially before the peak of SM $\bar{\nu}_e$ fluxes, to reduce SM condemnations. This is evidenced by the improved limits on $g$ from \cref{fig:sensitivity} for $m_\phi \lesssim \unit[5]{MeV}$ and $g\gtrsim 2\times 10^{-11}$ , where $\delta t \lesssim \unit[0.1]{s}$ for typical values of $E_\nu$.

It is important to note that there are currently uncertainties from different modeling of SN explosions as well as discrepancies between SN 1987A neutrino data and modeling, see e.g. Ref.~\cite{Li:2023ulf, Fiorillo:2023frv}.
These uncertainties will be significantly reduced after a future supernova.
For example, Hyper-Kamiokande can distinguish different SN modeling in high precision \cite{Hyper-Kamiokande:2021frf}.
Nevertheless, we have tested the robustness of our exclusion limits, the results of which we show in Supplemental Material.

Since it is unlikely that we will fully understand SN dynamics even with this abundance of new information, it is critical to have complementary methods for probing BSM scenarios in order to verify any potential observations.
This work constitutes a complementary method.
We make the code used in this work publicly available on \href{https://github.com/jlazar17/NuTel_SNe_BSM/}{GitHub} so that the impact of SN modeling uncertainties on the sensitivity of neutrino telescopes to BSM scenarios can be further explored.


\textbf{\textit{Acknowledgements}}---We want to thank Thomas Janka and Daniel Kresse for providing the data from Garching supernovae simulations---available at \url{https://wwwmpa.mpa-garching.mpg.de/ccsnarchive/} in machine-readable form---and also for stimulating discussions.
We also thank Jackapan Pairin for their help designing figures in this work.
YYL is supported by the NSF of China through Grant No. 12305107, 12247103.
CAA are supported by the Faculty of Arts and Sciences of Harvard University, the National Science Foundation (NSF), the NSF AI Institute for Artificial Intelligence and Fundamental Interactions, the Research Corporation for Science Advancement, 
and the David \& Lucile Packard Foundation.
CAA and JL were partially supported by the Alfred P. Sloan Foundation for part of this work.
JL is supported by the Belgian American Educational Foundation.


\bibliography{ic_sn_hnl}

\appendix




\appendix

\ifx \standalonesupplemental\undefined
\setcounter{page}{1}

\setcounter{figure}{0}

\setcounter{table}{0}
\setcounter{equation}{0}
\fi
\renewcommand{\thepage}{Supplemental Methods and Tables -- S\arabic{page}}

\renewcommand{\figurename}{SUPPL. FIG.}

\renewcommand{\tablename}{SUPPL. TABLE}
\renewcommand{\theequation}{A\arabic{equation}}

\newcounter{SIfig}
\renewcommand{\theSIfig}{SUPPL. FIG. \arabic{SIfig} }

\subsection{Sensitivity to Dipole Magnetic Moment Portal}
\label{app:magnet_moment}

To demonstrate the broad applicability of the technique proposed in this Letter, we demonstrate the calculation for another BSM model, the active-to-sterile neutrino transition magnetic moment described in Refs.~\cite{Magill:2018jla,Brdar:2020quo,Kamp:2022bpt,Brdar:2023tmi}.
In this model, the SM Lagrangian is extended to include:
\begin{align}
    \mathcal{L} \supset \sum_\alpha d_\alpha \bar{N}\sigma_{\mu\nu} \nu^{\alpha} F^{\mu\nu}-\frac{M_N}{2} \bar{N}^c N + \text{h.c.}\,,
    \label{eq:Lag}
\end{align}
where $\nu^{\alpha}$ and $N$ represent active and sterile neutrinos, respectively.
Further, $F^{\mu\nu}$ is the field strength tensor of the electromagnetic field, $d_\alpha$ is the dimensionful coefficient of this dimension-5 term, and $M_N$ is sterile neutrino mass.

We assume flavor universal interaction, i.e., $d_\alpha \equiv d$.
We consider the two production channels for $N$ inside the SN: $\nu e^- \to N e^-$ and $\nu \gamma \to N$.
Both processes occur due to the interaction term in \cref{eq:Lag}; after $N$ are produced, they decay to active neutrinos and photons which is again realized through the same term in the Lagrangian, with the decay width for $N\to\nu\gamma$ given by $\Gamma_N = 6d^2 M_N^3/4 \pi$~\cite{Plestid:2020vqf}.

\begin{figure}[b]
    \centering
    \includegraphics[width=0.47\textwidth]{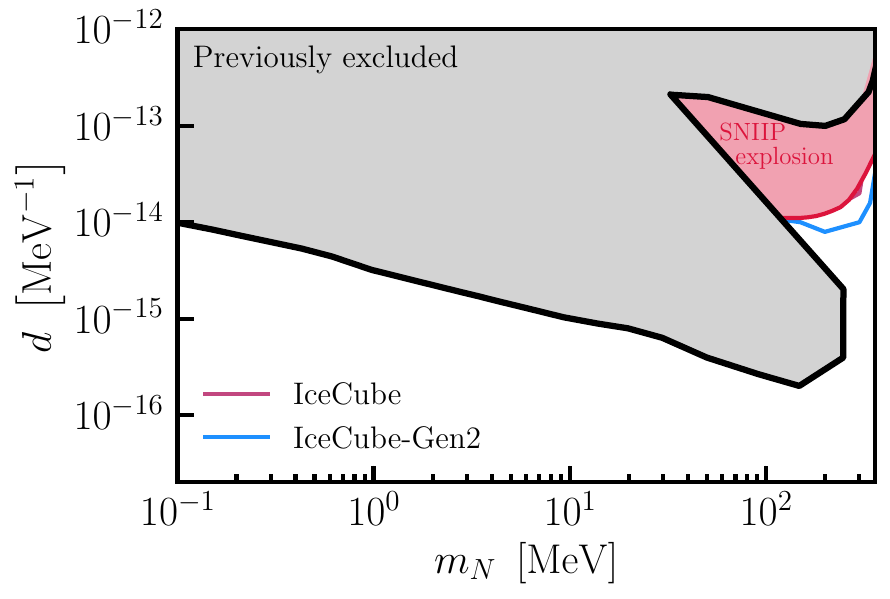}
    \caption{\textbf{\textit{Exclusion sensitivities for the magnetic moment case.}}
    The lines on this plot show the exclusion sensitivity for IceCube, IceCube-Gen2.
    The shaded regions are excluded by the combination of energy loss requirement, non-observation of photon and neutrino signals from SN1987A studied in~\cite{Brdar:2023tmi} and by constraints on the energy release from SN explosion (SNIIP explosion)~\cite{PhysRevLett.128.221103, Chauhan:2024nfa}.
    }
    \label{fig:magnetic_moment_sensitivity}
\end{figure}

In Suppl.~\cref{fig:magnetic_moment_sensitivity} we show the constraints attainable with the same method in the main text, together with regions previously excluded by energy loss, non-observations of photons and neutrinos from SN1987A~\cite{Brdar:2023tmi}, and constraints on the energy release from low-energy SN explosions (SNIIP)~\cite{PhysRevLett.128.221103, Chauhan:2024nfa}.
The additional phase space this technique is able to constrain below that already ruled out by the SNIIP results is relatively small.
Note that by neglecting the $N$ production channel $\nu p^+\to N p^+$ inside SN, the limits we obtained are conservative relative to the SNIIP region taken from~\cite{Chauhan:2024nfa}.

\subsection{Normalization Mismodeling}
\label{app:normalization}

We make the first effort to assess the robustness of our method to supernovae flux mismodelling.
From Ref.~\cite{Li:2023ulf}, we estimated the theoretical uncertainty on the SM flux normalization to be approximately 20\%.
Thus, we injected a SM signal scaled by 1.2 and 0.8 to capture the $1\sigma$ deviation of the flux from the nominal calculation.
We then used the unscaled template to draw pseudodata including the shape parameter $\mu^{\rm BSM}_i$ and performed the fit, test-statistic construction, and sensitivity calculation as described in the main text.
We found that the fitting procedure recovered the injected mismodeling coefficient, and consequently, the test-statistic distributions and exclusion sensitivity remained unchanged. Thus, we conclude that our procedure is robust to theoretical uncertainties on the normalization of the SM flux, assuming the shape differences are well-constrained.
This is a reasonable assumption given previous results~\cite{Li:2023ulf}.

\subsection{False Positive and Negative from Mismodelling}
\label{app:false_positive}

Another worrying possibility is that incomplete knowledge of the PNS parameters or incorrectly picked discrete modeling choices may lead to false positives or negatives with this technique.
\begin{figure}[t!]
    \centering
    \includegraphics[width=0.47\textwidth]{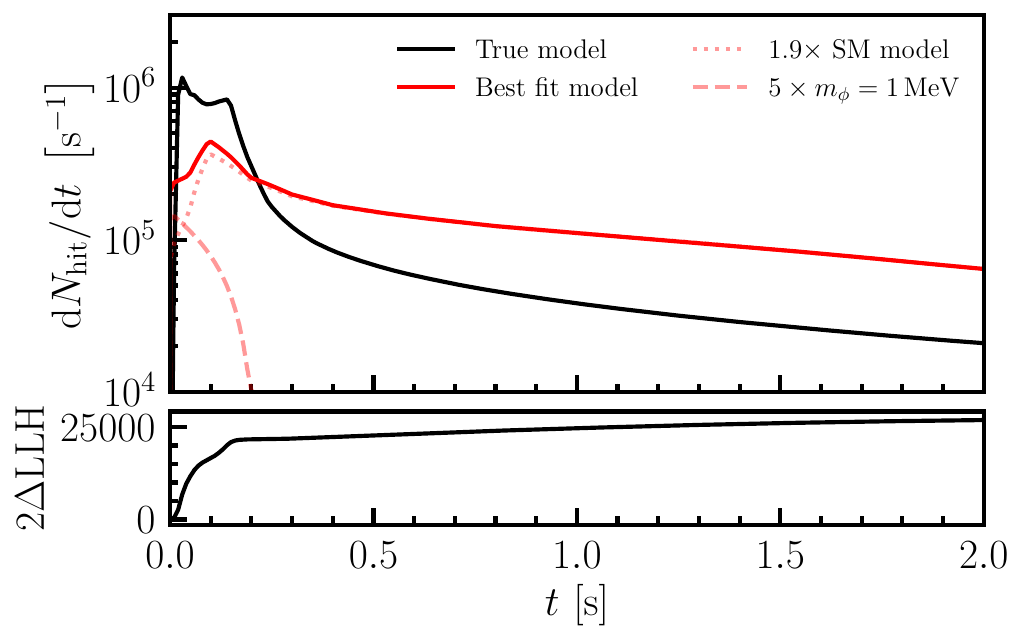}
    \caption{\textbf{\textit{False positive rejection.}}
    In the top panel, the black line shows the expected hit distribution from which the pseudodata was generated.
    Respectively, the dotted and dashed red lines show the best fit of the baseline SM and BSM models scaled by the median best-fit normalizations, and the solid red line shows the sum of both contributions.
    In the bottom panel, we show the median of  test statistic between the injected alternative SM-only model and the SM-plus-BSM model.
    The large value of the test statistic indicates an extremely strong rejection of the alternative SM-plus-BSM model.
    }
    \label{fig:false_positive}
\end{figure}
To test for the potential for a false positive, we computed the expected SM distribution of events in IceCube for a PNS of $13m_{\odot}$~\cite{Nakazato:2012qf}.
We then drew pseudodata from this alternative SM-only model and fit two different models: the injected model, i.e., the alternative SM-only model from which the pseudodata was generated,
and the baseline SM-plus-BSM.
We carried out this procedure with the nominal SM template and the BSM templates just above and just below our sensitivity for each $m_\phi$.
In Suppl.~\cref{fig:false_positive}, we compare the injected model and the median best-fit values for the baseline SM-plus-BSM that were the least strongly rejected.
This model had $m_{\phi}=1\,\mathrm{MeV}$ and $g_{\phi}=10^{-10}$.
Even in this case---which was most compatible with the injected data---we find that the alternative SM-only model was preferred by more than 25,000 test-statistic units.
This indicates an extremely strong rejection of the alternative SM-only model.

\begin{figure}[b]
    \centering
    \includegraphics[width=0.47\textwidth]{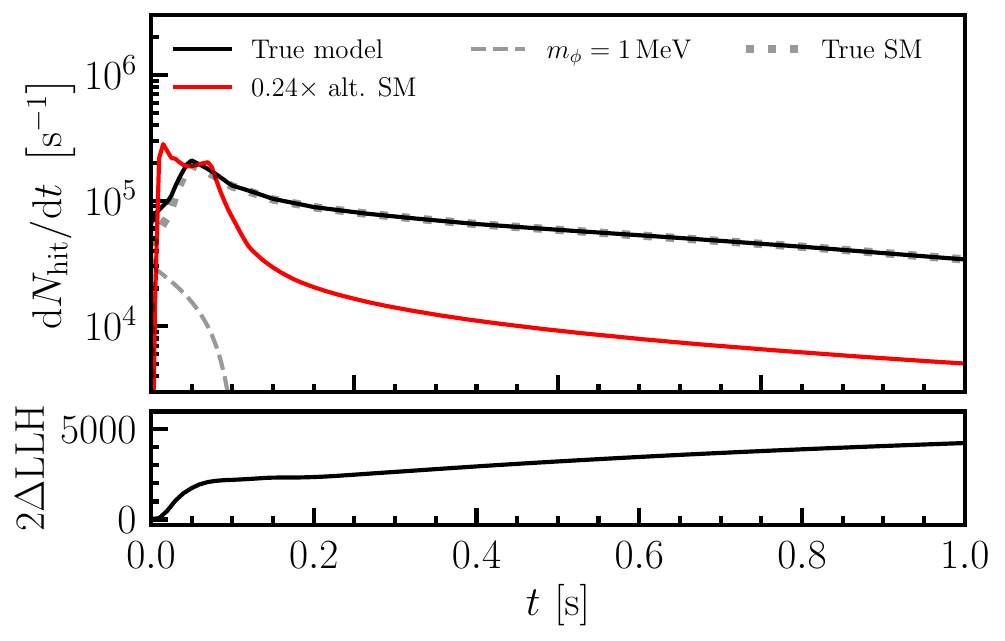}
    \caption{\textbf{\textit{False negative rejection.}}
    In the top panel, the black line shows the expected hit distribution for the injected SM-plus-BSM model with $m_{\phi}=1\,\mathrm{MeV}$ and $g=10^{-10}$, with the contribution from  SM and BSM fluxes represented by the dashed and dotted lines respectively.
    The red line shows the alternative SM flux scaled by the median of the best-fit normalizations.
    The bottom panel shows the median of test statistic between the best fit for true SM-plus-BSM model and the alternative SM model.
    }
    \label{fig:false_negative}
\end{figure}
To test the potential for a false negative, we perform a similar procedure; however, in this case, we drew pseudodata from our baseline SM-plus-BSM and then fit the alternative SM-only and the baseline SM-plus-BSM.
As before, we carried out this procedure with the BSM models just above and just below our sensitivity at each mass point.
We found that our procedure strongly rejected the alternative SM model in favor of the injected SM-plus-BSM model.
The model that was the least strongly rejected---which still was rejected by more than 4,000 test statistic units---was the same is in the false positive modeling.
In Suppl.~\cref{fig:false_negative}, we show this case.

Together, these findings indicate that our fitting procedure is robust against uncertainties on the mismodeling of the SM SN flux.
Furthermore, they indicate that this technique will not produce spurious positive or negative results.
This technique can also be adopted to select the best fit of the data among several SN modeling choices, as is common in particle physics when a smooth parameterization of the simulation is not available or is computationally not tractable.

\subsection{SN Pinching Parameter Uncertainty}
\label{sec:shape_uncertainty}

Despite the wealth of data the next SN will bring, there will likely still be uncertainties on certain aspects.
In particular, the exact shape of the SN spectrum, typically captured in the so-called pinching parameter, is expected to have uncertainties on the order of 10\% \cite{Rosso_2018}.
To test the robustness of our procedure against these uncertainties, we produced altered SM SN neutrino fluxes by changing the pinching parameter coherently across all times by $\pm10\%$.
We then used these new SN neutrino fluxes to generate new BSM fluxes for representative low- and high-mass cases ($g_\phi = 10^{-9.25}$ and $g_\phi = 10^{-11.8}$ for $m_\phi = \unit[0.1]{MeV}$ and $m_\phi = \unit[150]{MeV}$, respectively).
These represent the regime where a late- and early-time BSM signals alter the standard light curve.

\begin{figure}
    \centering
    \includegraphics[width=0.9\linewidth]{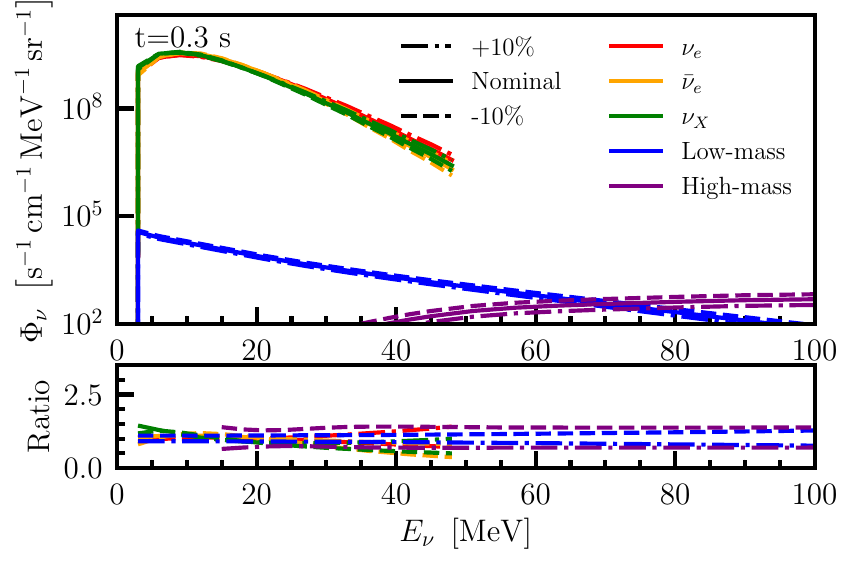}
    \caption{\textbf{\textit{Flux comparison between nominal and pinched cases}}
    The top panel shows the per-flavor flux of neutrinos in each scenario.
    The bottom panel shows the ratio of the modified fluxes to the nominal flux.
    Note the relatively flat behavior of the ratio as a function of energy for both the low- and high-mass cases.
    The ratio is constant across time, resulting in a nearly normalization-only change to the BSM light curves.
    }
    \label{fig:pinch_compare}
\end{figure}

While the SM fluxes had moderate shape changes, the differences in the BSM fluxes were almost energy- and time-independent, with normalization shifts between 7\% and 36\%; see \cref{fig:pinch_compare}.
Thus, since our analysis fits the normalization, we expect it to be robust against these changes.
To test this, we injected the varied SM and BSM neutrino fluxes and fit using the nominal values.
We found that the background test-statistic distribution did not change and still followed a modified $\chi^{2}$ distribution.
The signal-plus-background test-statistic distribution shifted to higher (lower) values for variations with higher (lower) overall normalizations.
In the low-mass (high-mass) case, the median value of these distributions was 4.80/2.85/7.53 (3.82/2.14/6.78) for the nominal, downward, and upward variations, respectively.
While these shifts are significant, we found that shifting the coupling by 25\% in the nominal case resulted in larger shifts to the median test statistic.
Thus, marginalizing over pinching parameter uncertainties should change the excluded coupling by less than 25\%.


\end{document}